\documentstyle[12pt]{article}

\catcode`@=11

\def\beqra{\begin{eqnarray}} \def\eeqra{\end{eqnarray}}
\def\beqast{\begin{eqnarray*}} \def\eeqast{\end{eqnarray*}}
\def\beq{\begin{equation}}      \def\eeq{\end{equation}}
\def\be{\begin{enumerate}}   \def\ee{\end{enumerate}}

\def\fo{\hbox{{1}\kern-.25em\hbox{l}}}
\def\fnote#1#2{\begingroup\def\thefootnote{#1}\footnote{#2}\addtocounter
{footnote}{-1}\endgroup}

\def\sppt{Research supported in part by the
Robert A. Welch Foundation and NSF Grant PHY 9009850}

\def\utgp{Theory Group\\ Department of Physics \\ University of Texas
\\ Austin, Texas 78712}

\def\ul{\underline}

\def\ch{\@startsection{section}{1}{\z@}{-3ex plus-1ex minus-.2ex}%
        {2ex plus.2ex}{\large\sc}}

\def\raisenot{\raise .5mm\hbox{/}}
\def\nota{\ \hbox{{$a$}\kern-.49em\hbox{/}}}
\def\notA{\hbox{{$A$}\kern-.54em\hbox{\raisenot}}}
\def\notb{\ \hbox{{$b$}\kern-.47em\hbox{/}}}
\def\notB{\ \hbox{{$B$}\kern-.60em\hbox{\raisenot}}}
\def\notc{\ \hbox{{$c$}\kern-.45em\hbox{/}}}
\def\notd{\ \hbox{{$d$}\kern-.53em\hbox{/}}}
\def\notbd{\ \hbox{{$D$}\kern-.61em\hbox{\raisenot}}} 
\def\note{\ \hbox{{$e$}\kern-.47em\hbox{/}}}
\def\notk{\ \hbox{{$k$}\kern-.51em\hbox{/}}}
\def\notp{\ \hbox{{$p$}\kern-.43em\hbox{/}}}
\def\notq{\ \hbox{{$q$}\kern-.47em\hbox{/}}}
\def\notW{\ \hbox{{$W$}\kern-.75em\hbox{\raisenot}}}
\def\notz{\ \hbox{{$Z$}\kern-.61em\hbox{\raisenot}}}

\def\notpa{\hbox{{$\partial$}\kern-.54em\hbox{\raisenot}}}

\def\7#1#2{\mathop{\null#2}\limits^{#1}}        
\def\5#1#2{\mathop{\null#2}\limits_{#1}}        

\def\inbar{\vrule height1.5ex width.4pt depth0pt}
\def\IB{\relax{\rm I\kern-.18em B}}
\def\IC{\relax\leavevmode\hbox{\,$\inbar\kern-.3em{\rm C}$}}
\def\ID{\relax{\rm I\kern-.18em D}}
\def\IE{\relax{\rm I\kern-.18em E}}
\def\IF{\relax{\rm I\kern-.18em F}}
\def\IG{\relax\leavevmode\hbox{\,$\inbar\kern-.3em{\rm G}$}}
\def\IH{\relax{\rm I\kern-.18em H}}
\def\II{\relax{\rm I\kern-.18em I}}
\def\IK{\relax{\rm I\kern-.18em K}}
\def\IL{\relax{\rm I\kern-.18em L}}
\def\IM{\relax{\rm I\kern-.18em M}}
\def\IN{\relax{\rm I\kern-.18em N}}
\def\IO{\relax\leavevmode\hbox{\,$\inbar\kern-.3em{\rm O}$}}
\def\IP{\relax{\rm I\kern-.18em P}}
\def\IQ{\relax\leavevmode\hbox{\,$\inbar\kern-.3em{\rm Q}$}}
\def\IR{\relax{\rm I\kern-.18em R}}
\def\sed{\hbox{{\sf S}\kern-.4em\hbox{\sf S}}}
\def\ZZ{\relax{\sf Z\kern-.4em Z}}
\def\smIR{\hbox{{\footnotesize\rm I}\kern-.2em\hbox{\footnotesize\rm R}}}
\def\smIO{\ \hbox{{\footnotesize\rm I}\kern-.4em\hbox{\footnotesize\bf O}}}
\def\smIQ{\ \hbox{{\footnotesize\rm I}\kern-.5em\hbox{\footnotesize\bf Q}}}
\def\IGa{\relax{\rm I}\kern-.18em\Gamma}
\def\IPi{\relax{\rm I}\kern-.18em\Pi}
\def\IQt{\relax\leavevmode\hbox{$\kern.3em\inbar\kern-.3em\Theta$}}
\def\IOm{\relax\hbox{$\kern3.48pt\inbar\kern1.8pt\inbar\kern-5.28pt\Omega$}}

\def\ca#1{\relax\ifmmode {\cal#1} \else$\cal#1$\fi}     
\def\Sf#1{\relax\ifmmode\hbox{\sf#1}\else{\sf#1}\fi}    
\def\fibby{\ifcase\@ptsize                      
                \font\tenrm=cmfib8\or           
                \font\elvrm=cmfib8 scaled\magstephalf\or        
                \font\twlrm=cmfib8 scaled\magstep1 \fi}         
\def\TeXey{\ifcase\@ptsize\or\or                
                \font\twlrm=cmr10 scaled\magstep1       
                \font\twlmi=cmmi10 scaled\magstep1      
                \font\twlit=cmti10 scaled\magstep1      
                \font\twlbf=cmbx10 scaled\magstep1\fi}  

\def\ch{\@startsection{section}{1}{\z@}{-3ex plus-1ex minus-.2ex}%
        {2ex plus.2ex}{\large\sc}}
\def\sch{\@startsection{subsection}{2}{\z@}{-1.5ex plus-1ex minus-.2ex}%
        {1pt plus.2ex}{\sc}}
\def\ssch{\@startsection{subsubsection}{3}{\z@}{-1ex plus-1ex minus-.2ex}%
        {1pt plus.2ex}{\small\sc}}
\def\seceq{\@addtoreset{equation}{section}
        \def\theequation{\thesection.\arabic{equation}}}        

\def\con{\ifmmode \hbox{\bf*} \else{\bf*}\fi}   
\def\scon{\ifmmode \hbox{\footnotesize\rm\bf*} \else{\footnotesize\rm\bf*}\fi}

\def\0#1{\relax\ifmmode\mathaccent"7017{#1}
        \else\accent23#1\relax\fi}              

\def\place#1#2#3{\vbox to0pt{\kern-\parskip\kern-7pt
                             \kern-#2truein\hbox{\kern#1truein #3}
                             \vss}\nointerlineskip}

\def\illustration #1 by #2 (#3){\vbox to #2{\hrule width #1 height 0pt depth
0pt
                                       \vfill\special{illustration #3}}}

\def\scaledillustration #1 by #2 (#3 scaled #4){{\dimen0=#1 \dimen1=#2
           \divide\dimen0 by 1000 \multiply\dimen0 by #4
            \divide\dimen1 by 1000 \multiply\dimen1 by #4
            \illustration \dimen0 by \dimen1 (#3 scaled #4)}}

\catcode`@=12

\begin{document}

\hfill{UTTG-17-92}

\hfill{July 1992}

\vfill
\vspace{24pt}
\begin{center}

{\bf Trivial Spectrum of Free 1+1 Light-Cone Strings\fnote{*}{\sppt}}

\vspace{24pt}

Eric Smith

\vspace{12pt}
\utgp
\vspace{36pt}

\ul{ABSTRACT}
\end{center}
\baselineskip=24pt

The BRST cohomology of 1+1 strings in a free light-cone gauge contains
only the two-dimensional tachyon, and excludes all excited states of
both matter and ghosts, including the special states that arise in
the continuum conformal gauge quantization and in the $c = 1$ matrix
models.  This exclusion takes place at a very basic level, and
therefore may signal some serious problems or at least unresolved
issues involved in this gauge choice.

\vfill
\pagebreak
\setcounter{page}{1}

Fundamental strings used to be defined by a requirement that they
possess only transverse physical excitations~\cite{oldstrings}.
That changed with the discovery that Polyakov's critical string theory
in $1+1$ dimensions possesses excited states of timelike oscillators as
nontrivial representatives of its BRST cohomology at special values of
the center of mass momentum~\cite{discstates}.  Since then these states
have been extensively studied~\cite{specstates,MinicnYang}, and while their
algebra is now well described~\cite{groundring}, their physical
interpretation remains obscure.  It would be useful to have a description
of these states in an easily generalizable light-cone gauge, to understand
both how they were missed by earlier quantizations and whether they have
any relevance in higher dimensional or effective string theories with
similar backgrounds.

An attempt to formulate such a description was begun in~\cite{self}.
It centered on a gauge which, in $1+1$ dimensions, leaves a free
theory of its unfixed degrees of freedom, but possesses propagating
ghosts and a nontrivial BRST charge.  This attempt appears to fail,
because unlike BRST quantization in conformal gauge, which admits many
states to the mass shell but excludes almost all the excited ones from a
reduced cohomology, the free light-cone gauge seems to exclude all excited
states directly from the mass shell.  This drastic exclusion takes place
at a very basic level, and could signal some serious inconsistency of this
gauge or of the implementation of it given in~\cite{self}.  Because
the program may have value, but the details remain problematic,
this note describes the calculation of the spectrum.

The following results are brought forward from previous work:  The
starting point is a Liouville-type action
\beq
{S}_{C} = {\frac{-1}{4 \pi \alpha '}}
          \int {{d}^{2}\sigma } \sqrt{g}
          \left( {g}^{ab}{\partial }_{a} {X}^{\mu }
                         {\partial }_{b} {X}^{\nu } {\eta}_{\mu \nu }
                 +\alpha ' n \cdot X {R}^{(2)} \right)
     					\label{eq:begaction}
\eeq
with world-sheet coordinates $(\tau ,\sigma )$ and metric ${g}_{ab}$
describing a Euclidean cylinder of spatial $(\sigma )$ period
$-2\sqrt{2}\pi $\footnote{This range is chosen so that the corresponding
complex coordinates take the usual form.},
and target space light-cone coordinates ${X}^{\pm } \equiv
{\left( {X}^{0} \pm {X}^{1}\right) }/\sqrt{2}$ with flat target space
metric ${\eta }_{\mu \nu }$ and $n$ a constant target space vector.
$\alpha '$ is the inverse string tension.
Fadeev-Popov~\cite{FnP} imposition of the gauge condition
\beq
{\hat g}^{--} \equiv \sqrt{g}{g}^{--} = 0; \;\;\;
\sqrt{g} = 1; \;\;\;
{\hat{X}}^{+} \equiv
{X}^{+} + \frac{\alpha '{n}^{+}}{2} \log {\sqrt{g}} = \tau
     					\label{eq:gaugecond}
\eeq
leaves the free actions
\begin{eqnarray}
{S}_{C+M} & = & \frac{1}{\pi } \int {{d}^{2}\sigma }\, {\hat g}^{++}
                \left[ \left( 1-\frac{{\partial }_{+}}{M} \right)
     \left( \frac{\sqrt{2}}{4\alpha '}{{\hat X}^{-},}_{+} \right)
    +\frac{1}{4\alpha '} \right] \nonumber \\
{S}_{GH}  & = & \frac{1}{\pi } \int {{d}^{2}\sigma }
                \left( w{u,}_{+} + z{v,}_{+} \right)
     					\label{eq:freeaction}
\end{eqnarray}
for matter and ghosts, respectively.  Here
${\hat g}^{++} \equiv \sqrt{g} {g}^{++}$,
${\hat X}^{-} = {X}^{-} - \frac{\alpha '{n}^{+}}{2} \log \sqrt{g}
         \equiv {X}^{-} - \frac{1}{\sqrt{2}M}       \log \sqrt{g}$,
and metric components and derivatives are indexed with respect to
world sheet coordinates ${\sigma }^{\pm} \equiv
{\left( \tau \pm \sigma \right) }/\sqrt{2}$.  The ghost (antighost)
components $u$,$v$,($w$,$z$) come from a decomposition of the
reparametrization ghosts ${c}^{+}$ and ${c}^{-}$:
\beq
{c}^{-} = u; \;\;\;\;
{c}^{+} = -(1+\frac{{\partial}_{-}}{M}) u
        + {e}^{-\sqrt{2}M\tau } v
     					\label{eq:ghdecomp}
\eeq
and ${\hat X}^{+}$-- and ${\hat g}^{--}$-- antighosts, named
respectively ${b}_{f}$ and ${\hat b}_{--}$:
\beq
{b}_{f}       = -M {e}^{\sqrt{2}M\tau } z; \;\;\;\;
{\hat b}_{--} = \frac{w}{2} -\frac{1}{2M} {e}^{\sqrt{2}M\tau } {z,}_{-}.
     					\label{eq:aghdecomp}
\eeq
The Weyl ghost is trivial and has already been integrated out.  The
subscript ${\scriptstyle C+M}$ on the matter action indicates that the
factor ${\hat g}^{++}/{4\alpha '}$ is a term extracted from the measure
and added to the canonical action to give the resulting full matter
action the symmetries of the usual Liouville theory\cite{DDKnJoe}.

The equations of motion in the matter sector are solved by a field
decomposition of the form\footnote{The additive constants which appear
here are a convenience which simplified the expression of the conserved
currents in~\cite{self}.}
\beq
{\hat g}^{++} = f + {e}^{-\sqrt{2}M\tau } g -2; \;\;\;\;
\frac{\sqrt{2}}{4\alpha '}{\hat X}^{-}
              = h - {e}^{\sqrt{2}M\tau } j
                  - \frac{\sqrt{2}}{4\alpha '} \tau,
     					\label{eq:mdecomp}
\eeq
with $f,g,h,j$ as well as $u,v,w,z$ functions only of ${\sigma }^{-}$
on shell.  Boundary conditions for the cylinder require periodicity of
all of these functions when
$\sigma \rightarrow \sigma - 2\sqrt{2}\pi $.
At this point it is convenient to take $\sigma = -is$ so that
${\sigma }^{-}$ becomes the complex variable w, in terms of which the
functions $f,g,h,j,u,v,w,z$ are analytic.  Because the currents of
interest are all tensors, it is convenient to map the cylinder to a
plane with coordinates ${\rm z} = {e}^{\rm w}$.  Then the measure for path
integration in this theory is specified by imposing the free field
propagators
\beq
  \left< vz' \right> = \left< uw' \right>
= \left< hf' \right> = \left< jg' \right> =
\frac{1}{{\rm z} - {\rm z}'}.
     					\label{eq:freeprop}
\eeq

The reparametrization symmetries for matter and ghosts have no simple
form in terms of fields $\theta u$, $\theta v$ (drawn from the ghost
decomposition~(\ref{eq:ghdecomp}) by adding a Grassmann parameter $\theta $
to the fields $u$,$v$), but the associated conserved currents take a
simple form on shell:
\beq
\begin{array}[b]{lcl}
{J}_{uC}  & = & \left( -f\,\partial h-\partial g\,j-2g\,\partial j
                       +\frac{2}{M}{\partial }^{2}h \right) ;\\
{J}_{vC}  & = & \left( mfj+2\,\partial j \right) ;\\
{J}_{uGH} & = & \left( -v\,\partial z+u\,\partial w+2\,\partial u\,w
     						    \right) ;\\
{J}_{vGH} & = & \left( u\,\partial z \right) .
\end{array}
     					\label{eq:securrents}
\eeq

The ${J}_{u}$s generate an analytic Virasoro algebra, and the
${J}_{v}$s an analytic $U(1)$.  There are no antianalytic currents
associated with these symmetries.  The BRST charge $Q$ takes the
standard form:
\beq
Q = \frac{1}{2\pi i}\oint {\rm dz}
  \left[ u{J}_{uC} + v{J}_{vC} + \frac{1}{2}
         \left( u{J}_{uGH} + v{J}_{vGH} \right) \right] .
     					\label{eq:qbrst}
\eeq

The conformal dimensions of the analytic fields are computed from
their operator product expansions with the currents ${J}_{u}$, and
take values ($1$,$2$,$0$,$-1$; $-1$,$1$,$2$,$0$) respectively for
($f$,$g$,$h$,$j$;$u$,$v$,$w$,$z$).  The
currents ${J}_{uC}$,${J}_{uGH}$ have dimension $2$ and central charge
$+28$,$-28$ respectively, and the currents ${J}_{vC}$,${J}_{vGH}$ have
dimension zero and no anomaly terms.  Of the OPEs of ${J}_{u}$ with
the free fields, the only anomalous product is ${J}_{uC}f'$, which
contains the term \( {4/{\left( {\rm z}-{\rm z}' \right) }^{3}M} \).

Decomposition of the analytic fields into oscillators is done in the
usual way, with
\beq
{\alpha }_{n} \equiv \frac{1}{2\pi i} \oint {\rm dz}\,
                     {\rm z}^{n+d-1} \alpha ({\rm z}); \;\;\;\;
\alpha ({\rm z}) = \sum_{n} \frac{{\alpha }_{n}}{{\rm z}^{n+d}},
     					\label{eq:oscdef}
\eeq
where $\alpha $ represents the name of the field and $d$ is its
conformal dimension.  The special names
${L}_{n},{K}_{n};{\lambda }_{n},{\kappa }_{n}$ will be given to the
expansion coefficients in
${J}_{uC},{J}_{vC};{J}_{uGH},{J}_{vGH}$, respectively.

The first point to notice about this theory is that the matter fields,
and not just their derivatives, decompose directly into analytic
fields and constants.  The consequences of this decomposition become
apparent when the generators of the current algebra are written in
terms of oscillators, as:
\begin{eqnarray}
{L}_{m}        & = & \sum_{n}
\left[ (m-n):{f}_{n}{h}_{m-n}: + (2m-n):{g}_{n}{j}_{m-n}: \right]
     							\nonumber \\
& & +\frac{2}{M} m(m+1){h}_{m} + {\delta }_{m}\nonumber \\
{\lambda }_{m} & = & \sum_{n}
\left[ (m-n):{v}_{n}{z}_{m-n}: + (-m-n):{u}_{n}{w}_{m-n}: \right]
                           - {\delta }_{m}\nonumber \\
{K}_{m}        & = & \sum_{n}
\left[ M{f}_{n}{j}_{m-n} \right] -2(m-1){j}_{m}\nonumber \\
{\kappa }_{m}  & = & \sum_{n}
\left[ n{z}_{n}{u}_{m-n} \right]
     					\label{eq:genosc}
\end{eqnarray}
and
\begin{eqnarray}
Q & = & \sum_{m} \left[ {u}_{m}{L}_{-m} + {v}_{m}{K}_{-m} \right]
     							\nonumber \\
 & &  + \sum_{m,n} \left[ \frac{(m-n)}{2} :{u}_{m}{u}_{n}{w}_{-m-n}:
                      +n {u}_{-m-n}:{v}_{m}{z}_{n}: \right]
  - {u}_{0},
     					\label{eq:brstosc}
\end{eqnarray}
where $:\; \; :$ denotes creation-annihilation normal ordering, and
the ordering constants in ${L}_{0},{\lambda }_{0}$ and $Q$ are fixed
by the action of the commutators on the ground state, as implied by the
current-current OPEs~\cite{Joebook}.

A standard technique for computing BRST cohomology is to write $Q$ as
a sum
\beq
Q = {u}_{0}\left( {L}_{0}+{\lambda }_{0} \right)
  + {v}_{0}\left( {K}_{0}+{\kappa  }_{0} \right)
  -2{w}_{0}\sum_{n>0} n{u}_{-n}{u}_{n} + {\hat Q}
     					\label{eq:qdecomp}
\eeq
and show how the cohomology of $Q$ relates to that of ${\hat Q}$ and
to the mass shell conditions that arise as a consequence of
reparametrization invariance~\cite{Joebook}.

The ghost zero mode sectors of this theory are slightly more
complicated than those for a simple Virasoro algebra, but it is still
straightforward to show that every BRST cohomology class can be
represented by a state $\left| {\rm phys} \right> $ for which
\beq
\left( {L}_{0}+{\lambda }_{0} \right) \left| {\rm phys} \right> = 0;
\;\;\;\;
\left( {K}_{0}+{\kappa  }_{0} \right) \left| {\rm phys} \right> = 0.
     					\label{eq:massshell}
\eeq

The elementary commutation relations implied by~(\ref{eq:freeprop})
make it possible to define number operators for the excitations
which annihilate the ground state and satisfy
\beq
\left[ {N}_{{\alpha }_{n}}, {{\alpha }_{-m}}^{k} \right]
= k{{\alpha }_{-m}}^{k}{\delta }_{n-m}; \;\;\;\; m>0 ,
     					\label{eq:numbercom}
\eeq
in terms of which
\begin{eqnarray}
{L}_{0}        & = & \sum_{n>0}
n \left[ {N}_{{f}_{n}}+{N}_{{g}_{n}}+{N}_{{h}_{n}}+{N}_{{j}_{n}} \right]
+1 \nonumber \\
{\lambda }_{0} & = & \sum_{n>0}
n \left[ {N}_{{u}_{n}}+{N}_{{v}_{n}}+{N}_{{w}_{n}}+{N}_{{z}_{n}} \right]
-1 \nonumber \\
{K}_{0}        & = & \sum_{n>0}
M \left[ {f}_{-n}{j}_{n}+{j}_{-n}{f}_{n} \right]
+\left( M{f}_{0}+2 \right) {j}_{0} \nonumber \\
{\kappa }_{0}  & = & \sum_{n>0}
-n \left[ {z}_{-n}{u}_{n}+{u}_{-n}{z}_{n} \right] .
     					\label{eq:gennum}
\end{eqnarray}

It follows immediately that all physical states must be products of
matter and ghost ground states, and that they satisfy the matter
zero-mode projection
\beq
\left( M{f}_{0}+2 \right) {j}_{0} \left| {\rm phys} \right> = 0.
     					\label{eq:momconst}
\eeq
Up to an additive constant ambiguity that arises from the addition of
total derivatives linear in ${\hat g}^{++}$ or ${\hat X}^{-}$ to the
action, ${f}_{0}$ and $-{e}^{\sqrt{2}M\tau }{j}_{0}$ are just the zero
frequency components of the momenta conjugate to
$\sqrt{2}{\hat X}^{-}/4\alpha '$ and ${\hat g}^{++}$, respectively.
The condition~(\ref{eq:momconst}) therefore has the same form as the
projection onto the tachyon state in the conformal gauge, though here
the generator of translations in ${\hat X}^{+}$ is tied through the
gauge condition to the choice of world-sheet coordinates and so to the
generator of translations in ${\hat g}^{++}$.

One possibility is that this is actually the correct spectrum of a
consistent $1+1$ string theory, and that the special states are somehow
artifacts of other methods of quantization.  However, the fact that
they are needed to obtain a consistent factorization of scattering
amplitudes in~\cite{MinicnYang}, as well as their presence in the
matrix models, makes this seem unlikely.  The alternative is that
there is some flaw or subtlety associated with this quantum theory
which needs yet to be explained.

I again am indebted to J. Polchinski for useful discussions and for
the use of his manuscript in preparation.  I also thank R. Rudd for
discussions.  This research was supported in part by the Robert A.
Welch Foundation, NSF Grant PHY 9009850, and the Texas Advanced
Research Program.

\pagebreak


\baselineskip=12pt

\begin{thebibliography}{20}

\bibitem{oldstrings}
P. Goddard and C. Thorn, Phys. Rev. Lett. B40 (1972) 235;\\
J.L. Gervais and B. Sakita, Nucl. Phys. B34 (1971) 477;\\
C.G. Callan, D. Friedan, E.J. Martinec and M.J. Perry, Nucl. Phys.
B262 (1985) 593;\\
R.C. Myers, Phys. Lett. B199 (1987) 371
\bibitem{discstates}
A.M. Polyakov, Mod. Phys. Lett. A2 (1987) 893;\\
V.G. Knizhnik, A.M. Polyakov, and A.B. Zamolodchikov, Mod. Phys. Lett.
A3 (1988) 819
\bibitem{specstates}
S. Mukherji, S. Mukhi and A. Sen, Phys. Lett. B266 (1991) 337;\\
U.H.Danielsson and D.J. Gross, Nucl. Phys. B366 (1991) 3;\\
A.M. Polyakov, Princeton preprint PUPT-1289, Lectures given at the
Jerusalem winter school (1991);\\
I.R. Klebanov and A.M. Polyakov, Mod. Phys. Lett. A6 (1991) 3273
\bibitem{MinicnYang}
D. Minic and Z. Yang, Phys. Lett. B274 (1992) 27
\bibitem{groundring}
E. Witten, Nucl. Phys. B373 (1992) 187;\\
D. Kutasov, E. Martinec and N. Seiberg, Phys. Lett. B276 (1992) 437
\bibitem{self}
E. Smith, U.T. Austin preprint UTTG-36-91, to appear in Nucl. Phys. B
\bibitem{FnP}
L.D. Faddeev and V.N. Popov, Phys. Lett. B25 (1967) 30
\bibitem{DDKnJoe}
F. David, Mod. Phys. Lett. A3 (1988) 1651;\\
J. Distler and H. Kawai, Nucl. Phys. B321 (1989) 509;\\
J. Polchinski, Nucl. Phys. B324 (1989) 123
\bibitem{Joebook}The author learned these methods from a manuscript on
String Theory and Conformal Field Theory by J. Polchinski, in
preparation.

\end{thebibliography}
\end{document}